\documentclass[
    ,final            
  ,numberedheadings 
  ]
  {aipproc}

\usepackage{multirow}

\layoutstyle{6x9}

\begin{document}

\title{Structure properties of even-even actinides}



\author{J.-P. Delaroche}{
  address={CEA/DAM Ile de France, DPTA/Service de Physique Nucl\'eaire, \\
  BP 12, 91680 Bruy\`eres-le-Ch\^atel, France}
}

\author{M. Girod}{
  address={CEA/DAM Ile de France, DPTA/Service de Physique Nucl\'eaire, \\
  BP 12, 91680 Bruy\`eres-le-Ch\^atel, France}
}
\author{\underline{H. Goutte}}{
  address={CEA/DAM Ile de France, DPTA/Service de Physique Nucl\'eaire, \\
  BP 12, 91680 Bruy\`eres-le-Ch\^atel, France}
}
\author{J. Libert}{
  address={Institut de Physique Nucl\'eaire, CNRS-IN2P3 and Universit\'e Paris XI \\
  15 rue Cl\'emenceau, 91406 Orsay, France}
}

\begin{abstract}
Structure properties of fifty five even-even actinides have been calculated using the Gogny D1S force
and the Hartree-Fock-Bogoliubov approach as well as the configuration mixing method. Theoretical results are
compared with experimental data.

\end{abstract}

\maketitle

\section{Introduction}
The existence of superheavy elements (SHEs) is a major issue in nuclear physics and
many experimental and theoretical efforts have been put in that direction during the last
decades. In order to achieve reliable predictions in the superheavy mass region, a good understanding
of the properties of heavy mass nuclei near and at the limits of stability
is essential.
In a recent work we have performed large scale microscopic calculations of structure properties of fifty five
even-even actinides at normal and isomeric potential deformations, namely $^{226-236}$Th, $^{228-242}$U, $^{232-246}$Pu,
$^{238-250}$Cm, $^{238-256}$Cf, $^{242-258}$Fm and$^{250-262}$No~\cite{Dela06}. Results on potential energy surfaces, barriers,
multipole moments, moments of inertia, shape isomers and positive parity phonon excitations, long-lived isomers, shape isomer half-lives have been discussed.
Calculations have been performed using the Gogny D1S force
together with the constrained Hartree-Fock-Bogoliubov (HFB) method as well as
blocking, and cranking HFB approaches and the configuration mixing method. Half-lives have been determined using the semi-classical WKB approximation.
In the present paper, we recall some of the results.

\section{Fission barriers}
\begin{figure}[h]
\includegraphics[height=8cm]{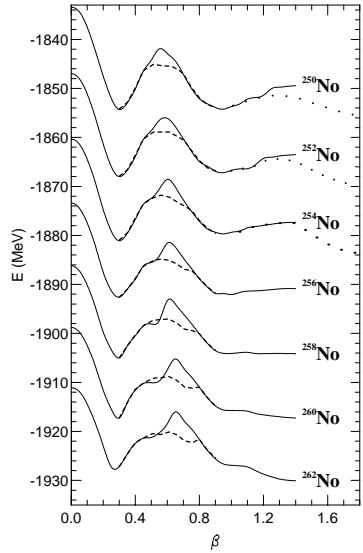}
\caption{Potential energy curves as function of $\beta$ for No isotopes. Thin solid lines are for axial barriers; dashed lines for triaxial inner barriers and dotted lines correspond to mass asymmetric outer barriers.}
\label{fig1}
\end{figure}

First, we present results from constrained-Hartree-Fock-Bogoliubov calculations using the Gogny D1S force and a sole constraint on the axial quadrupole deformation $\beta$. This deformation parameter is related to the quadrupole moments $q_{20}$ and $q_{22}$ through $\beta =\sqrt{\frac{\pi}{5}}\ \frac{\sqrt{q_{20}^{2}+3q_{22}^{2}}}{A\ <r^{2}>}$,
with
$<r^{2}> \, = \frac{3}{5}\,\left( r_{0}\,A^{1/3}\right)^{2}$,
and $r_{0}\,=\,1.2$ fm. In these calculations, at low elongation the triaxial degree of freedom has been left free, whereas parity has been broken at large elongation. On Fig.~\ref{fig1} are plotted the potential energy curves for Nobelium isotopes as functions of $\beta$.
Solid lines correspond to axial shapes, dashed lines to triaxial ones and dotted ones to asymmetric shapes.The main features are: i) triaxial inner barriers are systematically lowered by up to 4 MeV when compared to the axial ones, ii) the outer barrier is found to be asymmetric for systems with N < 152 and symmetric for more neutron-rich systems, and finally iii) super deformed minima appear to be washed out for N > 156.
These features are illustrated here in the case of Nobelium isotopes, but they are common features of all the seven studied isotopic chains.

\section{Shape isomers}
\begin{figure}[h]
\includegraphics[scale=0.4,angle=-90]{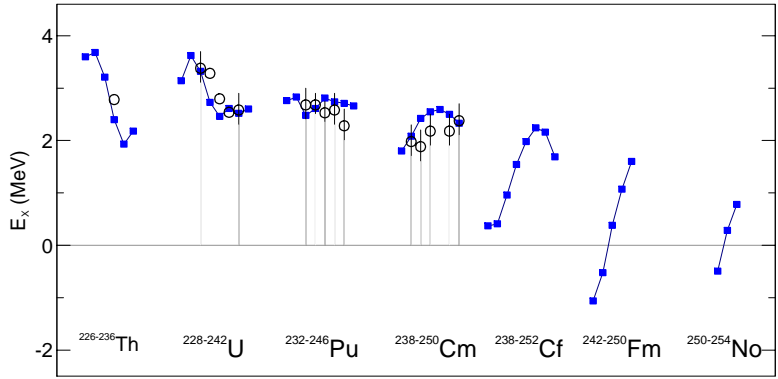}
\caption{Excitation energy (in MeV) of the shape isomers expressed with respect to normal deformed ground states for Th to No isotopes.}
\label{fig2}
\end{figure}

Positive parity collective states have been determined using the Generator Coordinate Method and the Gaussian Overlap Approximation. 
A Bohr Hamiltonian-like equation is solved for the five collective quadrupole coordinates, that is for axial and triaxial moments as well as the three Euler angles.
Results for shape isomers are shown in Fig.~\ref{fig2}.
A good agreement between predictions (full symbols) and experimental data (open symbols) is found in Th, U, Pu and Cm isotopes~\cite{NucDatSheet},~\cite{Bjorn80}. We predict a global lowering of isomer energies as A increases. Superdeformed states are even found to be lower in energy than normal deformed states in $^{242,244}$Fm and $^{250}$No. Nevertheless, as these states are only a few hundred keV below the octupole unstable outer barrier, they may not survive as bound states.

Partial $\gamma$-back decay and fission half-lives of all the shape-isomers have been determined using the WKB method. The collective masses introduced in the calculations have been evaluated using the Adiabatic Time Dependent Hartree-Fock formalism.
We have found that: i) shape isomers in Th and U decay by $\gamma$-emission, ii) fission and $\gamma$-back decay are competing for Pu and Cm, and finally iii) shape isomers in Cf, Cm and No decay through fission.
\begin{figure}[h]
\includegraphics[scale=0.4,angle=-90]{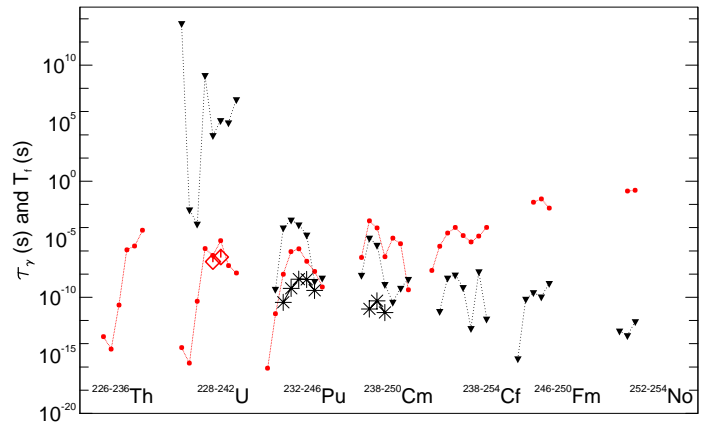}
\caption{Partial half-lives for $\gamma$-back and fission decays of shape isomers. Crosses ($^{236,238}$U) and stars ($^{236 - 244}$Pu, $^{240 - 244}$Cm)
are experimental data for $\gamma$-back and fission decays,
respectively. Dots and triangles are one-dimension WKB
predictions for $\gamma$-back and fission decays, respectively.}
\label{dureedevie1D}
\end{figure}

Furthermore, longer half-lives are predicted for N = 146 Th, U, Pu and Cm isotopes, which suggests that N = 146 is a magic number at super deformation.

\section{Inner barrier heights}
\begin{figure}[h]
\includegraphics[scale=0.4,angle=-90]{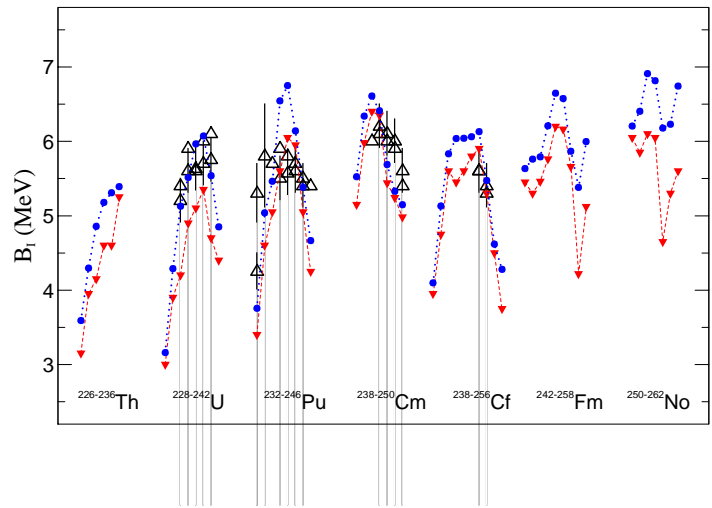}
\caption{Theoretical inner barriers heights for Th to No isotopes obtained using method A (dots) and method B (triangles) compared to experimental data (open symbols).}
\label{barr}
\end{figure}

Inner fission barrier heights have been determined using two different methods. In the first method (method A), the barrier is calculated as the energy difference between the top of the triaxial HFB inner potential (corrected by the zero point energy) and the energy of the collective normal-deformed lowest energy $0^+$ state. In the second method (method B) the barrier corresponds to the energy difference between the last collective state located in the first well and the collective normal deformed lowest energy state. Theoretical results are plotted as dots (method A) and triangles (method B) for the different studied isotopes in Fig.~\ref{barr}. Predictions are compared to experimental data shown as open symbols.
A good agreement is obtained with experimental data. A bell-shape structure is predicted in all the seven isotopic chains and N = 146 isotopes are found to have the larger inner barrier heights. 

The difference observed between the two sets of calculations provides a measure of the theoretical uncertainty in the determination of the barrier heights. This uncertainty is of the order of 500 keV.

\section{Third well spectroscopy}

In some nuclei, namely $^{246}$Pu, $^{248,250}$Cm and $^{252,254}$Cf  isotopes, we predict a shallow potential minimum on top of the triaxial inner barrier.

\begin{figure}[h]
\includegraphics[scale=0.3,angle=0]{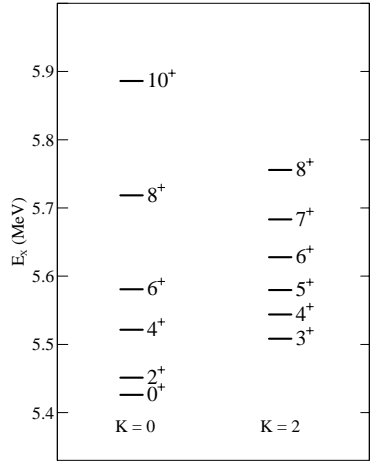}
\caption{Excitation energies of ${\pi\,=\,+}$ collective levels, and band structure at the $^{250}$Cm shallow
triaxial potential minimum.}
\label{cm250p3hbniv}
\end{figure}
Maxima of collective masses present in the vicinity of these minima provide stability of collective levels at such intermediate deformations, especially in $^{250}$Cm.
In Fig.~\ref{cm250p3hbniv} are plotted the quasi-rotational K = 0 and K = 2 bands
predicted at intermediate deformation in $^{250}$Cm.
The quasi-rotational bands do
not show regular structures typical of those predicted for a standard
triaxial rotor.  The present band features are related to the mean
deformations attached to the levels, whose values display some spread
over the third potential.
These levels located between 5 and 6 MeV, could qualitatively explain the  broad structures observed in fission probability measurements~\cite{Britt78}
at such energies in some heavy actinides with N $\simeq$ 154.

\section{conclusion}

A rich structure information has been collected over the years in experimental studies of actinides. We have used them
as playgrounds to systematically challenge mean-field based methods predictions through the actinide region, the gateway to superheavy actinides.
Only a few results have been presented here. More informations, for instance concerning spin isomers, phonon states and moments of inertia
can be found in~\cite{Dela06}.

\end{document}